\documentclass[11pt]{article}
\usepackage[utf8]{inputenc}
\pdfoutput=1
\usepackage{amssymb,amsmath,mathrsfs,enumerate}
\usepackage{graphicx,rotate,multicol}
\usepackage{float}
\usepackage{tocloft}
\usepackage{subfig}
\usepackage[margin=10pt,labelfont=bf]{caption}
\usepackage{cite}
\usepackage{soul}
\usepackage[T1]{fontenc}
\usepackage[utf8]{inputenc}
\usepackage[colorlinks=true,
            linkcolor=blue,
            urlcolor=blue,
            citecolor=teal]{hyperref}

\makeatletter
\let\Hy@linktoc\Hy@linktoc@page
\makeatother

\usepackage{color}
\definecolor{ourcolor}{rgb}{0.7, 0.25, 0.05}
\usepackage{tikz,braket}
\long\def\rpl#1!!#2!!{\textcolor{red}{#1} \textcolor{blue}{#2}}

\let\tilde=\widetilde
\let\hat=\widehat
\let\bar=\overline

\def \order(#1){{\mathcal O} \left(#1 \right)}

\evensidemargin=\oddsidemargin
\allowdisplaybreaks

\title{\color{black}{\bf Dark Energy from pNGB Mediated Dirac Neutrino Condensate}}

\author {\sf Ujjal Kumar Dey,$^{a,}$\footnote{ujjal@cts.iitkgp.ernet.in} 
\hspace{4pt}  Tirtha Sankar Ray,$^{a,b,}$\footnote{tirthasankar.ray@gmail.com}
\hspace{4pt}  Utpal Sarkar,$^{b,}$\footnote{utpal@phy.iitkgp.ernet.in} \\[10pt]
\small\em $^a$Centre for Theoretical Studies, 
		Indian Institute of Technology Kharagpur,
		Kharagpur 721302, India\\
\small\em $^b$Department of Physics, Indian Institute of Technology, Kharagpur 721302, India\\
}

\date{}

\begin{document}

\maketitle

\begin{abstract}
We consider an extension of the Standard Model that provide an unified description of eV scale neutrino mass and  dark energy. An explicit model is presented by augmenting the Standard Model with an $SU(2)_L$ doublet scalar, a singlet scalar and right handed neutrinos where all of them are assumed to be charged under a global $U(1)_X$ symmetry. A light pseudo-Nambu-Goldstone Boson, associated with the spontaneously broken $U(1)_{X}$ symmetry, acts as a mediator of an attractive force leading to a Dirac neutrino condensate, with large correlation length, and a non-zero gap  in the right range providing a cosmologically feasible dark energy scenario. The neutrino mass is generated through the usual Dirac seesaw mechanism. Parameter space, reproducing viable dark energy scenario while having neutrino mass in the right ballpark, is presented.
\end{abstract}

\newpage

\hrule \hrule
\tableofcontents
\vskip 10pt
\hrule \hrule 

%%%%%%%%%%%%%%%%%%%%%%%%%%%
\section{Introduction}
%%%%%%%%%%%%%%%%%%%%%%%%%%%%
The lion's share of the energy budget of the Universe is in the form of Dark Energy (DE), which unlike normal matter, produces a negative pressure driving the observed accelerated expansion of the Universe. Astrophysical observations reveal that in the present epoch the DE shares almost 69\% of the total energy in the Universe while visible and dark matter together make up to 31\%~\cite{Ade:2015xua}. Various models explaining the origin of DE have been proposed in the literature, for a brief review see e.g.,~\cite{Wetterich:1987fm, Ratra:1987rm, Copeland:2006wr, Bamba:2012cp, 2012IJMPD..2130002Y, 2013tasi.conf....1C}. On the other hand, the existence of  highly suppressed neutrino mass is a long standing issue with the Standard Model (SM) of particle physics. Although neutrino oscillation experiments can not measure the exact mass of the neutrinos, but only the mass differences between neutrinos of different generations, neutrinos are supposed to have masses in the eV or sub-eV range. The tantalizing closeness of the dark energy scale ($\sim$ meV) and the mass splitting of neutrinos hints at a possible connection between these two sectors. Scenarios of mass varying neutrinos where the mass of the neutrinos are dependent on the quintessence field called acceleron~\cite{Gu:2003er, Fardon:2003eh, Peccei:2004sz, Li:2004tq, Barger:2005mn, Brookfield:2005bz, Brookfield:2005td, Ringwald:2006ks, Barbieri:2005gj, Takahashi:2005kw, Fardon:2005wc, Ichiki:2008st, Ma:2006mr, Bhatt:2008hr, Franca:2009xp, Ghalsasi:2014mja, Geng:2015haa} have been proposed in the literature. Another possibility is considering the neutrino (or some other fermion) condensate  as the source of  dark energy~\cite{Antusch:2002xh, Kapusta:2004gi, Yajnik:2005pr, Bhatt:2009wb, Yajnik:2011mh, Dvali:2016uhn,  Inagaki:2016vkf}. 
In this context, it has been shown that Dirac-type neutrinos can show superfluidity providing a viable dark energy candidate. However, the effect is sizeable if the massive neutrinos couple to an unnaturally light scalar~\cite{Kapusta:2004gi}. This problem can be partially evaded when the neutrinos are assumed to be pseudo-Dirac in nature, and they form condensate via the interaction of a real singlet scalar field~\cite{Bhatt:2009wb}. The possibility of a Born-Infeld condensate coupled to neutrinos explaining small Dirac mass for neutrinos and dark energy is considered in~\cite{Addazi:2016oob}.
In this letter we propose a unique variant of the neutrino condensate proposal for the dark energy, in which Dirac neutrinos couple to a pseudo-Nambu-Goldstone Boson (pNGB), which gives an attractive force between the left-handed and the right handed components, forming a condensate. In earlier proposals for neutrino condensate with Dirac neutrinos~\cite{Kapusta:2004gi}, it was realized that the mediator scalar field has to be an $SU(2)_L$ doublet, and hence, the effective four neutrino interaction would be highly suppressed ($\propto f^2/m_{\rm Med}^2 \approx f^2/m_{\rm EW}^2$), making it cosmologically insignificant. One can relax this constraint by introducing a light singlet scalar field, but the mixing of the singlet with the doublet will bring in similar suppression factor, making the model useless. In our novel scheme to circumvent this problem, we introduce two new concepts. The mediator being a pNGB, the smallness of its mass is naturally explained and this explains why the dark energy dominates our universe only at the present time. The other virtue of this model is that the mixing between the singlet and the doublet is given by the ratios of the two vacuum expectation values (vevs), which are comparable and hence there are no suppression from the mixing. This is ensured by another interesting seesaw mechanism, where we start with two Higgs doublets, and one of them gets a very small vev, with Electroweak scale masses for all the components. 
The mediator is a pNGB that arises due to the spontaneous breaking of a $U(1)_{X}$ global symmetry. A soft breaking gives rise to a tiny mass of the pNGB which is technically natural~\cite{tHooft:1979rat} and further insulated from the UV because of the shift symmetry. We also take care of the neutrino mass by a Dirac seesaw like mechanism~\cite{Gu:2006dc}. Therefore in an unified setup we address the smallness of neutrino mass as well as the origin of DE. The weak scale phenomenology include possible charged scalar at the TeV scale that can provide complementary signatures of this framework.
The rest of the letter is organized as follows. In the next section we lay out the details of our model. In section~\ref{sec:nucond} we discuss the neutrino condensation formation. Allowed parameters and other numerical details are discussed in section~\ref{sec:cosm}. Finally we summarize and conclude in section~\ref{sec:concl}.

%%%%%%%%%%%%%%%%%%%%%%%%%%%%%%%%%%%%%%%
%%%%%%%%%%%%%%%%%%%%%%%%%%%%%%%%%%%%%%%
\section{The Model}
\label{sec:model}
\begin{figure}[!htbp]
\centering
\includegraphics[width=0.5\textwidth]{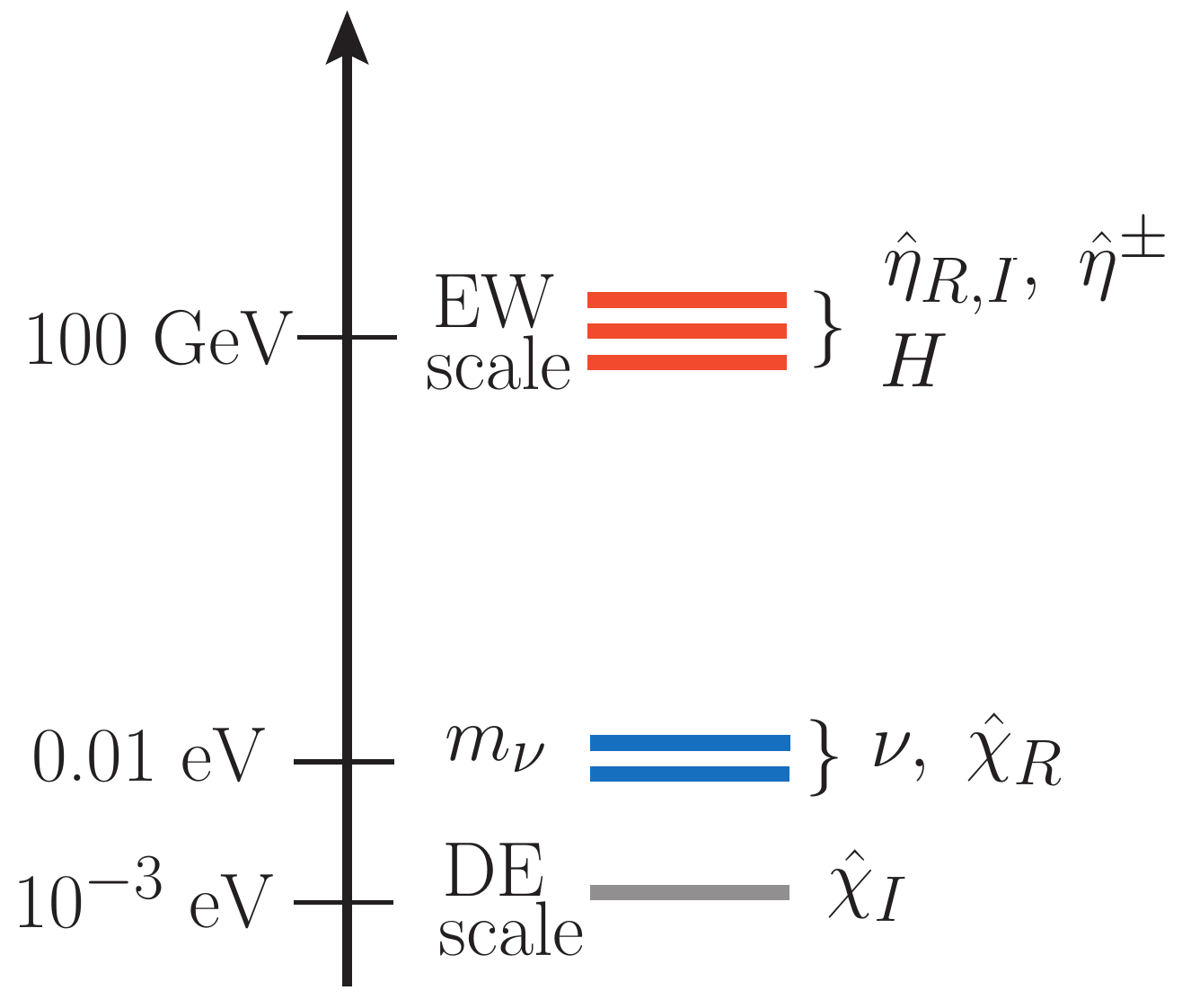}
\caption{Qualitative description of the mass spectrum of the model.}
\label{fig:massScale}
\end{figure}

%%%%%%%%%%%%%%%%%%%%%%%%%%%%%%%%%%%%%%%
%%%%%%%%%%%%%%%%%%%%%%%%%%%%%%%%%%%%%%%
We consider the following extension of the SM. Along with SM field content we add a right handed neutrino\footnote{In this discussion we only consider one generation of neutrinos for simplicity of discussion. Surely, three right handed neutrinos (RHN) will be required to obtain three massive neutrinos. However, our scenario can easily be generalized by adding two more RHNs.} $\nu_{R}$, (required for neutrino Dirac masses via a seesaw mechanism of the scalar doublets), one scalar doublet $\eta$ and one scalar singlet $\chi$.
\begin{table}[h]
\centering
\begin{tabular}{|c|c|c|c|c|}
\hline
            & $\Phi$          & $\eta$          & $\chi$ & $\nu_{R}$ \\ \hline 
\hline
$SU(2)_{L}$ & 2               & 2               & 1      & 1         \\ \hline
$U(1)_{Y}$  & $\frac{1}{2}$ & $\frac{1}{2}$ & 0      & 0         \\ \hline
$U(1)_{X}$  & 0               & 1               & 1   & 1      \\ \hline 
\end{tabular}
\caption{Field content and relevant quantum numbers. All other SM fields are $U(1)_{X}$ singlets.}
\label{tab:fields}
\end{table}
With these field content a Dirac mass to the neutrino should arise by the coupling of the the left-handed lepton doublet to the right handed singlet neutrino through the SM Higgs doublet $\Phi$. Given the GeV scale vev of the Higgs, this demands an unnaturally small Yukawa coupling to be in consonance with the observed tiny neutrino mass. Therefore, we introduce an $U(1)_{X}$ global symmetry to forbid such couplings. The relevant quantum numbers for various fields are shown in table~\ref{tab:fields}. With these quantum numbers the only invariant term in the Lagrangian connecting the left and right handed neutrinos is given by,
\begin{align}
\mathscr{L}_{\nu} = y \bar{L}\tilde{\eta} \nu_{R} + {\rm h.c.}\;,
\label{eq:Lnu}
\end{align} 
where $\tilde{\eta} = i\sigma_{2}\eta^{\ast}$. Within the usual Dirac seesaw mechanism a tiny vev for the $\eta$ can be arranged for by introducing  the following Lagrangian,
\begin{align}
\mathscr{L}_{\rm med} \supset ~ & m_{\eta}^{2}\eta^{\dagger}\eta + 
                        %\delta (\eta^{\dagger}\eta)^{2} %+ 
                        %\alpha (\Phi^{\dagger}\Phi) 
                         %  (\eta^{\dagger}\eta)
                       % + \beta (\Phi^{\dagger}\eta)
                         % (\eta^{\dagger}\Phi)  
                       % \notag \\ 
                       % & + \gamma (\eta^{\dagger}\eta) 
                        %    (\chi^{\ast}\chi) + 
                            \sqrt{2}\mu_{\eta} \left(
                        \chi \eta^{\dagger}\Phi + {\rm h.c.}
                        \right). 
\label{eq:LInt}                        
\end{align}
For simplicity we do not take into account the other possible quartic terms. Notice that a tadpole  for $\eta$ is generated when both $\Phi$ and $\chi$ gets a vev. This is arranged easily by  introducing the following potential for these fields,
\begin{align}
\mathscr{L}_{\rm vis} = -\mu_{\Phi}^{2}\Phi^{\dagger}\Phi 
                        + \lambda_{\Phi}(\Phi^{\dagger}\Phi)^{2} 
                         -\mu_{\chi}^{2}\chi^{\ast}\chi + 
                         \lambda_{\chi}(\chi^{\ast}\chi)^{2}
                         + \lambda_{\Phi\chi}\Phi^{\dagger}\Phi \chi^{\ast}\chi\;.
\label{eq:LPhi}                         
\end{align}  
The vev of $\eta$ is now suppressed by the $\eta$ mass given by the parameter $m_\eta$ in Eq.~\ref{eq:LInt}, which can be large. The quartic coupling between $\chi$ and $\Phi$ will be crucial to obtain a small vev for the field $\chi$. 
%Admittedly these couplings can lead to altered Higgs phenomenology leading to severe constraints \cite{Khachatryan:2016vau}, however we expect the coefficients can be tuned to ameliorate these issues without affecting the discussion here.
%
After the symmetry breaking $SU(2)_{L}\otimes U(1)_{Y}\otimes U(1)_{X} \to U(1)_{\rm em}$ the fields $\Phi$ and $\chi$ gets vevs $v_{\Phi}$ and $v_{\chi}$ respectively. While the Goldstones arising from the electroweak sector go into making the $W$ and $Z$ bosons massive by the usual SM Higgs mechanism, we are left with one massless Goldstone associated with the spontaneous breaking of the $U(1)_X$ global symmetry.
A small mass for the pNGB can be generated by introducing a soft breaking of the  global $U(1)_{X}$,
\begin{equation}
\mathscr{L}_{\not{U(1)_{X}}} = -M_{\chi}^{2}(\chi^{2} + {\rm h.c.}).
\label{eq:LBrk}
\end{equation}
We will assume the soft breaking parameter $M_\chi  \sim 10^{-3}~ \mbox{eV} \ll v_\Phi,$  is at the cosmological constant scale. This is technically natural, {\it \`{a} la} t'-Hooft~\cite{tHooft:1979rat}, as the vanishing $M_{\chi}$ enhances the symmetry of the theory.
A discussion of the  the spectrum arising from the Lagrangian in Eqs.~\ref{eq:Lnu}-\ref{eq:LBrk} is now in order. After the $U(1)_X$ breaking by the vev of $\chi$, the electroweak symmetry breaking by $\Phi$ would induce an eV scale vev to $\eta$. This can, in turn, give meV scale Dirac mass to the neutrino for $\mathcal{O}(1)$ Yukawa coupling. In this scenario, the smallness of the vev of $\eta$ is ensured by restricting the vev of $\chi$ and the mass of $\eta$. 
The interaction between the neutrinos can be mediated by the pNGB that is predominantly a combination of imaginary part of $\eta$ and $\chi$. However, for the pNGB, the mixing with the doublet is given by the ratio of vevs of $\eta$ and $\chi$, which will be arranged to be of order one within this framework. Since the mass of the imaginary part of $\chi$ is protected by symmetry, as it is a pNGB, the suppression factor in the effective four neutrino interaction is orders of magnitude larger ($\propto f^2/m_\chi^2 \approx f^2/m_\nu^2$) than the models of Dirac neutrino condensates studied so far.
In the component form we can write the scalars as,
\begin{gather}
\eta = \frac{1}{\sqrt{2}}
\begin{pmatrix}
\sqrt{2}\eta^{+}\\
v_{\eta} + \eta_{R} + i \eta_{I}
\end{pmatrix},~
\Phi = \frac{1}{\sqrt{2}}
\begin{pmatrix}
\sqrt{2}\phi^{+}\\
v_{\Phi} + h + iz
\end{pmatrix}, \\
\chi = \frac{1}{\sqrt{2}}(v_{\chi} + \chi_{R} + i \chi_{I})\;.
\end{gather} 
%Obviously we have the usual SM Higgs doublet $\Phi$,
%\begin{align}
%\Phi = \frac{1}{\sqrt{2}}
%\begin{pmatrix}
%\sqrt{2}\phi^{+}\\
%v_{\Phi} + h + iz
%\end{pmatrix}\;.
%\end{align}
The minimization of the relevant potential yields,
\begin{gather}
v_{\eta}^{2} = \frac{v_{\Phi}^{2}v_{\chi}^{2}\mu_{\eta}^{2}}
               {m_{\eta}^{4}}, \notag \\
v_{\Phi}^{2} = \frac{4\lambda_{\chi}\mu_{\Phi}^{2}-2ab}
                 {4\lambda_{\Phi}\lambda_{\chi}-b^{2}}, \quad
v_{\chi}^{2} = \frac{4a\lambda_{\Phi}-2b\mu_{\Phi}^{2}}{4\lambda_{\Phi}\lambda_{\chi}-b^{2}}\;,
\label{eq:vevSQ}
\end{gather}
where $a = (2M_{\chi}^{2} + \mu_{\chi}^{2})$ and $b = \lambda_{\Phi\chi}- 2(\mu_{\eta}/m_{\eta})^{2}$.
To produce a large mixing between the $ \eta_{I}$ and  $\chi_{I}$ we need $v_{\chi} \approx v_{\eta} \sim m_\nu \sim \mbox{eV}$. This can be easily arranged by setting $b = 0$. We will assume $m_{\eta}\sim \lambda_{\Phi\chi} v/\sqrt{2} $ and $\mu_{\eta}\sim \lambda_{\Phi\chi}^{2}v/2$.
In the basis $\{z, \eta_{I}, \chi_{I}\}$, the mass matrix squared can be written as
\begin{align}
\mathcal{M}_{C}^{2} = 
\begin{pmatrix}
\frac{v_{\chi}^{2}\mu_{\eta}^{2}}{m_{\eta}^{2}} & v_{\chi}\mu_{\eta} & 
-v_{\eta}\mu_{\eta} \\
v_{\chi}\mu_{\eta}  & m_{\eta}^{2}  & v_{\Phi}\mu_{\eta} \\
-v_{\eta}\mu_{\eta}  & v_{\Phi}\mu_{\eta}  & 4M_{\chi}^{2} + 
\frac{v_{\Phi}^{2}\mu_{\eta}^{2}}{m_{\eta}^{2}}
\end{pmatrix}\;,
\label{eq:MCSQ}
\end{align}
Note that the determinant of $\mathcal{M}_{C}^{2}$ is zero which implies that at least one of the eigenvalues is zero. This reflects the fact that one of the eigenstate will be the Goldstone of the SM $Z$ boson. The diagonalization of $\mathcal{M}_{C}^{2}$ will imply a rotation in the basis $\{z, \eta_{I}, \chi_{I}\}$ to the diagonal $\{\hat{z}, \hat{\eta}_{I}, \hat{\chi}_{I}\}$.
Now we can approximate various parameters in the model as,
\begin{subequations}
\begin{gather}
v_{\Phi} =\frac{\mu_{\Phi}}{\sqrt{\lambda_{\Phi}}} \equiv v \sim 100 ~{\rm GeV},
\\
v_{\chi} = \sqrt{\frac{\mu_{\chi}^{2} + 2M_{\chi}^{2}}{\lambda_{\chi}}} \equiv w \sim {\rm eV}; ~~ 
v_{\eta} \sim w,
\\
M_{\chi}  \sim 10^{-3}~{\rm eV}.
\end{gather}
\label{eq:vevapprox}
\end{subequations}
With these choices the three eigenvalues for $\mathcal{M}_{C}^{2}$, and hence the diagonal basis states, in terms of small quantity $w/v$ is given by,
\begin{align}
\hat{z} &\simeq -z; 
          ~~ m_{\hat{z}}^{2} = 0,  \\
\hat{\eta}_{I} &\simeq \frac{1}{\sqrt{2}}\left(
              \chi_{I} + \eta_{I}\right);
          ~~m_{\hat{\eta}_{I}}^{2} \simeq \lambda_{\Phi\chi}^{2}v^{2} + 
\mathcal{O}\left(w^{2}\right)
\\
\hat{\chi}_{I} &\simeq \frac{1}{\sqrt{2}} \left(
              \chi_{I} - \eta_{I}\right);
          ~~m_{\hat{\chi}_{I}}^{2} \simeq M_{\chi}^2
               \left(2 + \mathcal{O}\left(\frac{w^{2}}{v^{2}}\right)\right).
\end{align}
We now make the following identifications. The massless state $\hat{z}$ represents the Goldstone of the SM $Z$ Boson. We will see that the ultra-light state $\hat{\chi}_{I}$, which is a pseudo-Nambu-Goldstone boson (pNGB), will be responsible for the attractive potential between the Dirac neutrinos in our model. This ultimately gives rise to the superfluid neutrino condensate state providing a viable dark energy candidate. 
Also in the basis $\{h, \eta_{R}, \chi_{R}\}$, the mass matrix squared can be written as,
%\begin{widetext}
\begin{align}
\mathcal{M}_{R}^{2} = 
\begin{pmatrix}
2\mu_{\Phi}^{2} + 
  \frac{v_{\chi}^{2}\mu_{\eta}^{2}}{m_{\eta}^{2}} 
    & v_{\chi}\mu_{\eta} 
      & v_{\eta}\mu_{\eta} + \frac{2v_{\Phi}v_{\chi}\mu_{\eta}^{2}}{m_{\eta}^{2}}\\
v_{\chi}\mu_{\eta}  & m_{\eta}^{2}  &  v_{\Phi}\mu_{\eta}
   \\
v_{\eta}\mu_{\eta} + \frac{2v_{\Phi}v_{\chi}\mu_{\eta}^{2}}{m_{\eta}^{2}}  & v_{\Phi}\mu_{\eta}  & 
   2\mu_{\chi}^{2} + 4 M_{\chi}^{2} +  
    \frac{v_{\Phi}^{2}\mu_{\eta}^{2}}{m_{\eta}^{2}}      
\end{pmatrix},
\label{eq:MRSQ}
\end{align}
%\end{widetext}
Then in the diagonal basis $\{\hat{h}, \hat{\eta}_{R}, \hat{\chi}_{R}\}$ the states and their masses are given by,
\begin{align}
\hat{h} &\simeq h; 
          ~~ m_{\hat{h}}^{2} \simeq 2 v^{2} \\
\hat{\eta}_{R} &\simeq \frac{1}{\sqrt{2}}\left(
               \chi_{R} + \eta_{R}\right);
              ~ m_{\hat{\eta}_{R}}^{2} 
           \simeq \lambda_{\Phi\chi}^{2}v^{2} + 
             \mathcal{O}\left(w^{2}\right),
             \\
   \hat{\chi}_{R} &\simeq \frac{1}{\sqrt{2}} \left(
              \chi_{R} - \eta_{R}\right);
              ~ m_{\hat{\chi}_{R}}^{2} \simeq  w^{2}.
\end{align}
The state $\hat{h}$ can be identified with the SM Higgs, $H$.  There is an additional real scalar state at the neutrino mass scale  $m_{\hat{\chi}_{R}}^{2} \sim \mbox{eV}^2$. There will also be charged scalar states $\hat{\eta}^{\pm}$ whose masses are proportional to $m_{\eta}^{2}$.  In figure~\ref{fig:massScale} we qualitatively show the mass spectrum of the model.
Now we concentrate on the Lagrangian associated with neutrinos, $\mathscr{L}_{\nu}$ given in Eq.~(\ref{eq:Lnu}). In terms of the component fields this can be written as,
\begin{align}
\mathscr{L}_{\nu} = \frac{y v_{\eta}}{\sqrt{2}}
                      \bar{\nu}_{L}\nu_{R} + 
                      \frac{y}{\sqrt{2}}
                       \bar{\nu}_{L}\eta_{R}\nu_{R} - 
                       \frac{i y}{\sqrt{2}}
                       \bar{\nu}_{L}\eta_{I}\nu_{R} 
                        - y \bar{e}_{L}\eta^{-} \nu_{R} 
                        + {\rm h.c.} 
\end{align}
Clearly the first term is the Dirac mass term for the neutrinos and the mass is given by,
$
m_{\nu} = yv_{\eta}/\sqrt{2} = 
          \frac{yv_{\Phi}v_{\chi}\mu_{\eta}}{\sqrt{2}m_{\eta}^{2}}\;.
$
With the choice of various parameters $m_{\nu}$ can easily be fixed at values $\sim$ eV. The second and the third terms can be recast in terms of the mass basis fields as,
\begin{align}
\mathscr{L}_{\nu}^{\rm Yuk} \supset \frac{iy}{2}
               \bar{\nu}_{L}\hat{\chi}_{I}\nu_{R} - 
               \frac{iy}{2}\bar{\nu}_{L}\hat{\eta}_{I}\nu_{R} +
               \frac{1}{2}y
               \bar{\nu}_{L}\hat{\chi}_{R}\nu_{R}
               - \frac{1}{2}y
               \bar{\nu}_{L}\hat{\eta}_{R}\nu_{R} + {\rm h.c.}
\end{align} 
Now among these four terms only the first term will give rise to a reasonably strong attractive potential between the neutrinos owing to the lightness ($\sim 10^{-3}$~ eV) of the state $\hat{\chi}_{I}$. In the next section we will discuss the effective contact interaction mediated by $\hat{\chi}_{I}$ and the relevant prescription for the formation of neutrino condensate.  

%%%%%%%%%%%%%%%%%%%%%%%%%%%%%%%%%%%%%%%%%%%%%%%%%%%%%%%%%%%%%%
%%%%%%%%%%%%%%%%%%%%%%%%%%%%%%%%%%%%%%%%%%%%%%%%%%%%%%%%%%%%%%
\section{Neutrino Condensate Formation}
\label{sec:nucond}
%%%%%%%%%%%%%%%%%%%%%%%%%%%%%%%%%%%%%%%%%%%%%%%%%%%%%%%%%%%%%%
%%%%%%%%%%%%%%%%%%%%%%%%%%%%%%%%%%%%%%%%%%%%%%%%%%%%%%%%%%%%%%
In this section we will briefly discuss the neutrino condensate formation. Here we will follow the prescription of~\cite{Kapusta:2004gi}. For small momentum transfer the $\hat{\chi}_{I}$ exchange between the neutrinos can effectively result in a contact interaction, the Hamiltonian for which is given by,
\begin{align}
H_{I} = \mathcal{C} (\bar{\nu}\nu)(\bar{\nu}\nu)
      = \mathcal{C} (\bar{\nu}_{L}\nu_{R} + \bar{\nu}_{R}\nu_{L})
         (\bar{\nu}_{L}\nu_{R} + \bar{\nu}_{R}\nu_{L}),
\label{eq:HInt}
\end{align}
where the effective coupling strength $\mathcal{C} = y^{2}/m^{2}_{\hat{\chi}_{I}}$. Now, in the Dirac basis the neutrinos can be represented as,
\begin{gather}
\nu_{L} = P_{L}\nu = \frac12 (1-\gamma_{5})\nu 
                   = \frac{1}{\sqrt{2}}
                     \begin{pmatrix}
                     \psi_{L} \\
                     -\psi_{L}
                     \end{pmatrix}, \\
\nu_{R} = P_{R}\nu = \frac12 (1+\gamma_{5})\nu 
                   = \frac{1}{\sqrt{2}}
                     \begin{pmatrix}
                     \psi_{R} \\
                     \psi_{R}
                     \end{pmatrix},
\end{gather}
where $\psi_{L}$ and $\chi_{R}$ are two-component spinors. Thus in the two-component form the interaction Hamiltonian can be written as,
%\begin{align}
%H_{I} = \mathcal{C}&\left(
%        \psi_{L}^{a\dagger}\psi_{L}^{b\dagger}
%        \psi_{R}^{b}\psi_{R}^{a} + 
%        \psi_{R}^{a\dagger}\psi_{R}^{b\dagger}
%        \psi_{L}^{b}\psi_{L}^{a}  \right. \notag \\
%        &\qquad \left.
%       + 2\psi_{L}^{a\dagger}\psi_{R}^{b\dagger}
%        \psi_{L}^{b}\psi_{R}^{a}
%        \right),
%\label{eq:HIntExpanded}        
%\end{align}
\begin{align}
H_{I} = \mathcal{C}\left(
        \psi_{L}^{a\dagger}\psi_{L}^{b\dagger}
        \psi_{R}^{b}\psi_{R}^{a} + 
        \psi_{R}^{a\dagger}\psi_{R}^{b\dagger}
        \psi_{L}^{b}\psi_{L}^{a}
       + 2\psi_{L}^{a\dagger}\psi_{R}^{b\dagger}
        \psi_{L}^{b}\psi_{R}^{a}
        \right),
\label{eq:HIntExpanded}        
\end{align}
where the spinor indices $a$ and $b$ run from 1 to 2. According to the theory of superconductivity, to form a condensate there should exist an overall attractive interaction between the pairing particles, for details see~\cite{book:Fetter}. The $Z$ boson exchange gives rise to a repulsive interaction between two left handed neutrinos~\cite{Caldi:1999db}, while the condensate can be formed through a spin zero pairing of the left and right handed neutrino mediated by the scalar $\hat{\chi}_{I}$. So the appropriate condensate can be represented as $\langle\psi_{L}^{a}\psi_{R}^{b}\rangle = \epsilon^{ab}D$ where $\epsilon^{ab}$ is the usual Levi-Civita tensor and $D$ is related to the gap. Thus, in the mean field approximation the interaction Hamiltonian can be written as,
\begin{align}
H_{I} = -2\mathcal{C}\epsilon^{ab}
                  \left( D\psi_{L}^{a\dagger} 
                          \psi_{R}^{b\dagger} + 
                         D^{\ast}\psi_{L}^{b} 
                          \psi_{R}^{a} \right).
\end{align}  
Starting from this interaction Hamiltonian one can follow the standard Nambu-Gor'kov formalism~\cite{Gorkov:1958, Nambu:1960tm} to derive the gap equation. In our case this energy gap is given by,
\begin{align}
\Delta = \sqrt{\frac{2\Lambda}{m_{\nu}}}
         \left(3\pi^{2}n_{\nu}\right)^{1/3}
         e^{-x}.
\end{align} 
A non-zero value of $\Delta$ signifies the successful formation a condensate made up of quasiparticles which are linear combination of left handed and right handed neutrinos. In table~\ref{tab:param} we present characteristic values of $\Delta$ for various coherence lengths. Here $\Lambda$ is the cut-off scale which we identify as the scale of neutrino decoupling which is $\sim$ MeV~\cite{Bhatt:2009wb} and $x$ is a dimensionless quantity given by,
\begin{align}
x = \frac{4\pi^{2}}
     {\mathcal{C}m_{\nu}\left(3\pi^{2}n_{\nu}\right)^{1/3}},    
\end{align}
where $n_{\nu}$ is typical neutrino number density which is $\sim$ 110/c.c. The critical temperature ($T_{c}$) and the Pippard coherence length ($\xi$) are given as, 
\begin{align}
T_{c} \simeq 0.57\Delta, \quad
\xi = \frac{e^{x}}{\pi \sqrt{2\Lambda m_{\nu}}}.
\end{align}
The condensate dynamics is relevant when $\xi$ is comparable to the inter-particle spacing. This is possible when the interaction between the neutrinos are sufficiently strong. 
%To have viable dark energy features we need large coherence length. 

%%%%%%%%%%%%%%%%%%%%%%%%%%%%%%%%%%%%%%%%%%%%%%%%%%%%%%%%%%%%%%
%%%%%%%%%%%%%%%%%%%%%%%%%%%%%%%%%%%%%%%%%%%%%%%%%%%%%%%%%%%%%%
\section{Cosmological Parameters and Numerical Results}
\label{sec:cosm}
%%%%%%%%%%%%%%%%%%%%%%%%%%%%%%%%%%%%%%%%%%%%%%%%%%%%%%%%%%%%%%
%%%%%%%%%%%%%%%%%%%%%%%%%%%%%%%%%%%%%%%%%%%%%%%%%%%%%%%%%%%%%%
%
\begin{figure}[t]
\centering
\includegraphics[scale=1.0]{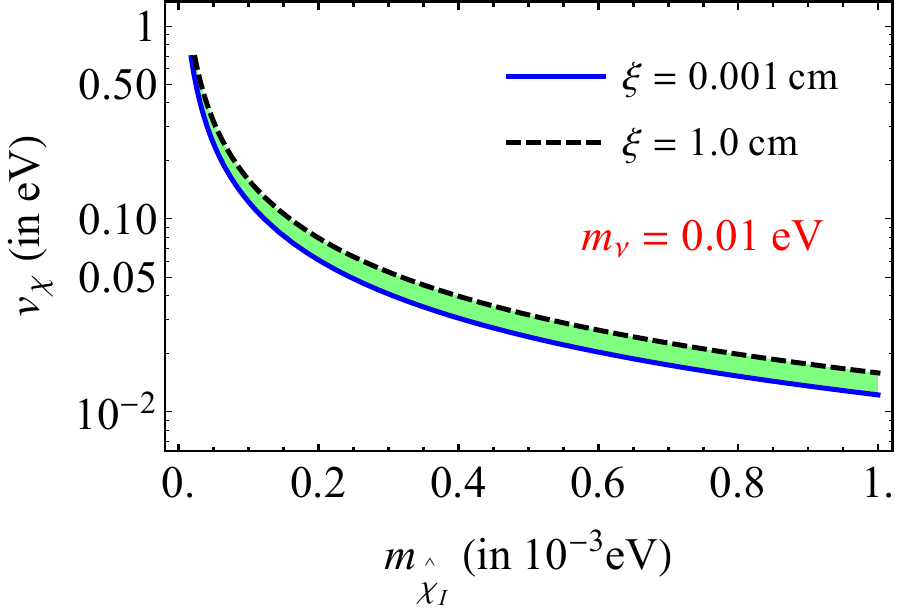}
\caption{Allowed values of $v_{\chi}$ and $m_{\hat{\chi}_{I}}$ for a characteristic neutrino mass $m_{\nu} = 0.01$ eV. The green shaded region corresponds to coherence length $\xi\in (0.001,1.0)$ cm for $y\lesssim 1$.}
\label{fig:mvfig}
\end{figure}
Now using the expressions for neutrino mass we can write the following relation,
\begin{align}
m_{\nu} = yv_{\eta}/\sqrt{2} = 
          \frac{yv_{\Phi}v_{\chi}\mu_{\eta}}{\sqrt{2}m_{\eta}^{2}} = 
          \sqrt{\frac{\mathcal{C}}{2}} 
\frac{m_{\hat{\chi}_{I}}v_{\Phi}v_{\chi}\mu_{\eta}}{m_{\eta}^{2}}\;.
\end{align}
Using the approximations mentioned in Eqs.~\ref{eq:vevapprox} we can recast this equation as,
%\begin{align}
$
m_{\hat{\chi}_{I}} v_{\chi} \simeq 
               \sqrt{\frac{2}{\mathcal{C}}}m_{\nu}.
$
%\end{align}
We are interested in the region where the Yukawa coupling $y$ remains perturbative. For such values of $y$ the main contribution to the condensate comes from the pNGB with mass $m_{\hat{\chi}_{I}} \sim 10^{-3}$ eV.
In figure~\ref{fig:mvfig} we show the allowed regions in $v_{\chi}$-$m_{\hat{\chi}_{I}}$ plane for coherence length $\xi\in (0.001,1.0)$ cm and characteristic neutrino mass of 0.01 eV. As can be seen from the figure for reasonable value of parameters in the theory one can reproduce the neutrino mass and at the same time have a non-zero  gap signifying formation of a condensate to explain the dark energy of the Universe without being effected by experimental constraints in quantum entanglement. Some other experimental consequences of neutrino condensates are discussed in~\cite{Azam:2010kw}. A brief discussion presented in~\cite{Bhatt:2009wb} shows that such a condensate behaves like a background scalar field which satisfy, in the absence of chemical potential, an equation of state $\omega = p/\rho = -1$.
\begin{table}[!htbp]
\centering
\begin{tabular}{|c|c|c|c|c|}
\hline
\begin{tabular}[c]{@{}c@{}}$\xi$\\ (cm)\end{tabular} & \begin{tabular}[c]{@{}c@{}}$\mathcal{C}$\\ (eV$^{-2}$)\end{tabular} & \begin{tabular}[c]{@{}c@{}}$\Delta$\\ (eV)\end{tabular} & \begin{tabular}[c]{@{}c@{}}$T_{c}$\\ (K)\end{tabular} & $x$    \\ \hline \hline
0.001                                                   & $1.3\times10^{6}$         & $1.8\times10^{-4}$                                         & $1.2$                                       & 10.0 \\ \hline
0.01                                                    & $1.1\times10^{6}$          & $1.8\times10^{-5}$                                         & $0.12$                                       & 12.3 \\ \hline
0.1                                                     & $9.2\times10^{5}$ & $1.8\times10^{-6}$                                         & $1.2\times10^{-2}$                                       & 14.6 \\ \hline
1                                                       & $7.9\times10^{5}$ & $1.8\times10^{-7}$                                         & $1.2\times10^{-3}$                                       & 16.9 \\ \hline 
\end{tabular}
\caption{Various parameters $\{\xi, \mathcal{C}, \Delta, T_{c}, x\}$ for characteristic $m_{\nu} = 0.01$ eV.}
\label{tab:param}
\end{table}
Some benchmark  values of the coherence length, energy gap, critical temperature and the parameter $x$ are shown in table~\ref{tab:param}. From this benchmark values we see that one can obtain a non-zero gap and $\mathcal{O}(1)$ K critical temperature for viable masses of neutrinos. This should be contrasted with~\cite{Kapusta:2004gi} where the critical temperature is many orders of magnitude smaller than 1 K for any neutrino mass precluding any cosmological implication. Also the present scenario is more relaxed in view of the neutrino mass and correlation length and critical temperature than the pseudo-Dirac case discussed in~\cite{Bhatt:2009wb}.
Thus, we find that the model provides the required amount of dark energy, as determined by the gap or the shift in background energy $\Delta$, for a reasonable value of the Pippard coherence length. The critical temperature, $T_c$, when the neutrino condensate starts dominating the universe initiating the observed acceleration, comes out to be close to the present background microwave radiation temperature, which explains another coincidence problem: why the universe started accelerating only now. In this scenario, the neutrino mass is related to the neutrino condensate and determined by the vev of the field $\eta$, which also explains why the amount of dark energy is comparable to the matter density of the universe.
%Since the pNGB is the background scalar field that determines the dynamics of the dark energy, the equation of state now becomes $\omega = p/\rho \approx -1$.
%

%
Within this setup we have a spectrum with  two  eV scale real scalars $\hat{\chi}_{R}$  and $\hat{\chi}_{I},$ that will affect big bang nucleosynthesis (BBN). These light states will contribute to the effective relativistic  degrees of freedom, $\delta N_{\rm eff} \approx 4/7 g^{\chi} (T_{\chi}/{T})^{4}\sim 1.14$, which is marginally allowed from BBN data \cite{Nollett:2014lwa}.  Further the charged Higgs, $\hat{\eta}^{\pm}$, arising from the second doublet can mediate the annihilation of right-handed neutrinos to the lepton pairs via $t$-channel processes modifying the BBN calculations. Following the prescription described in~\cite{Davidson:2009ha},   a typical $v_{\eta} = v_{\chi} \sim 0.5$ eV and $\mathcal{O}(1)$ Yukawa couplings  is consistent with the charged Higgs mass of order  $\sim$ 100 GeV as presented in Fig.~\ref{fig:mvfig}.

%%%%%%%%%%%%%%%%%%%%%%%%%%%%%%%%%%%%%%%
\section{Summary and Conclusion}
\label{sec:concl}
%%%%%%%%%%%%%%%%%%%%%%%%%%%%%%%%%%%%%%%
In this letter we have presented a simple scalar extension of the  SM that provides Dirac mass to the neutrino and at the same time explains the dark energy of the Universe through a  neutrino condensate mediated by a light pNGB scalar associated with the global abelian symmetry in the neutrino sector. Due to the attractive force mediated by the light pNGB the neutrino condensate forms and the resulting gap, $\Delta$, for a reasonable value of the coherence length determines the required amount of dark energy in the Universe. A single, technically natural small parameter, protected from UV physics by a shift symmetry, in the theory correlates the smallness of the neutrino mass and the light mass of the mediator. We find that in a region of parameter space the theory can accommodate the observed data for neutrino mass and dark energy parameters without large finetuning. 
%%%%%%%%%%%%%%%%%%%%%%%%%%%%%%%%%%%%%%%

%%%%%%%%%     Acknowledgements    %%%%%%%%%%%%%%%%%%%
\paragraph*{Acknowledgements\,:} We would like to thank Tarak Nath Maity, Sandip Bera, S. S. Mandal and A. Taraphder for useful discussions. UKD acknowledges the support from Department of Science and Technology (DST), Government of India under the fellowship reference number PDF/2016/001087 (SERB National Post-Doctoral Fellowship). TSR is partially supported by ISIRD Grant, IIT-Kharagpur. TSR also acknowledges the hospitality provided by ICTP, Italy,  under the Associate program, during the completion of the project. US acknowledges support from a research grant of DST, Government of India, associated with J. C. Bose Fellowship.

%%%%%%%%%%%%%%%%%   References %%%%%%%%%%%%%%%%%%%%%%%%%%%%%%%%%%%%
\bibliographystyle{JHEP}
\bibliography{ref.bib}

\end{document}